# Exceptional points in lossy media enable decay-free wave propagation

Alexander Yulaev[1,2,†], Sangsik Kim[1,3,†], Qing Li[1,4], Daron A. Westly[1], Brian J. Roxworthy[1,5],

Kartik Srinivasan[1], and Vladimir A. Aksyuk[1]*

[1]Physical Measurement Laboratory, National Institute of Standards and Technology,

Gaithersburg, MD 20899, USA.

[2]Department of Chemistry and Biochemistry, University of Maryland, College Park, MD 20742,

USA.

[3]Department of Electrical and Computer Engineering, Texas Tech University, Lubbock, TX

79409, USA.

[4]Department of Electrical and Computer Engineering, Carnegie Mellon University, Pittsburgh,

PA 15213, USA.

[5]Aeva, Inc. 555 Ellis Street, Mountain View, CA 94043, USA.

[†]These authors contributed equally to this work

*Corresponding author: vladimir.aksyuk@nist.gov

## Abstract

**Waves entering a spatially uniform lossy medium typically undergo exponential decay, arising from either the energy loss of the Beer-Lambert-Bouguer transmission law or the evanescent penetration during reflection. Recently, exceptional point singularities in non-Hermitian systems have been linked to unconventional wave propagation[1,2], such as the predicted extremely spatially broad constant-intensity**

**guided modes[3]. Despite such promises, the possibility of decay-free wave propagation in a purely lossy medium has been neither theoretically suggested nor experimentally realized until now. Here we discover and experimentally demonstrate decay-free wave propagation accompanied by a striking uniformly distributed energy loss across arbitrary thicknesses of a homogeneous periodically nanostructured waveguiding medium with exceptional points. Predicted by coupled-mode theory and supported by fully vectorial electromagnetic simulations, hundreds-of-waves deep penetration manifesting spatially constant radiation losses are experimentally observed in photonic slab waveguides. The uniform, decay-free radiative energy loss is measured across the entire structured waveguide region, regardless of its length. While the demonstrated constant-intensity radiation finds an immediate application for generating large, uniform and surface-normal free-space plane waves directly from the photonic chip surface, the uncovered decay-free wave phenomenon is universal and holds true across all domains supporting physical waves, opening new horizons for dispersion-engineered materials empowered by exceptional point physics.**

Idealized linear energy-conserving systems with time-reversal symmetry are Hermitian, described by a Hamiltonian with real eigenvalues, and the evolution of states controlled by a unitary operator. However, dissipation processes are ubiquitous in nature, breaking time-reversal symmetry and giving rise to Non-Hermitian systems. Non-Hermitian systems can manifest exceptional points (EPs), the singularities in parameter space where several eigenvalues and eigenvectors become degenerate. Systems with EPs have recently attracted attention due to their fundamentally different behavior from Hermitian or dissipative counterparts with non-degenerate solutions.

EPs are encountered in systems with cross-coupling between their modes comparable to the difference in loss or gain. In photonics, EPs give rise to a plethora of counterintuitive phenomena and enable new ways to control light generation and propagation[1,2] such as discriminating and controlling lasing modes[4-8], boosting the sensitivity of resonators to perturbations[9,10], loss-induced transparency[11], asymmetric power oscillations[12], non-reciprocal light transmission and propagation[13], unidirectional invisibility[14,15], and unique topological features[16-19]. Even though EPs were initially studied in systems with balanced combinations of gain and loss, later studies expanded to fully passive lossy systems with EP-containing spectra[11,20]. A variety of such non-Hermitian photonic systems have been reported, including ring/disc resonators[4-8,13], coupled waveguides[11,12,20], Bragg reflectors[21], and photonic crystals[14,22,23]. Under specific carefully tuned conditions, EPs are known to exist in band diagrams of infinite periodically structured wave media, such as photonic gratings[24]. Understanding wave propagation through such dispersion-engineered lossy materials is important for applications of optical metamaterials and integrated photonic devices involving coupling between photonic guided modes and free-space modes. Recently, it has been predicted that non-Hermitian optical systems with balanced gain and loss can result in a specific class of solutions with constant-intensity profiles[3], which are highly desirable but difficult to achieve experimentally, and has been verified only in macroscale acoustical waveguides with discrete gain-loss inclusions[25].

Here, we predict theoretically and demonstrate experimentally that a periodically structured passive medium tuned to operate at an exceptional point can manifest decay-free wave penetration with unusual, linearly varying wave amplitudes and spatially uniform energy losses over arbitrary long penetration distances. This wave propagation regime occurs whenever forward and backward traveling waves are coupled via a symmetric lossy intermediary mode (IM). It is experimentally realized in a large photonic slab waveguide nanostructured by a periodic etched grating carefully tuned to create EPs in the waveguide's band diagram, while coupling the guided waves to a collimated surface-normal free-space mode playing the IM role. When the frequency of an incident guided wave is tuned to the EP, the wave penetration pattern switches from a conventional quick ($\approx$ 40 wavelengths) exponential decay due to the Bragg reflection to a very deep penetration across the full length of the structured waveguide, as long as 400 waves, directly observable by the spatially constant-radiated intensity across the device. Coupled mode theory (CMT) and direct numerical modeling show that the amplitudes of the propagating and counterpropagating modes in this regime become linear as a function of distance, with their slopes defined by the length of the device, and their linear combination coupled to the radiative loss via the intermediate mode (IM) becomes spatially constant. The discovered wave propagation regime near EPs requires a few simple conditions and broadly applies across physical domains supporting propagating waves. Our photonic implementation finds immediate application in integrated photonics circuits requiring the generation of well-collimated top-hat free-space beams emanated vertically[26,27], for example, in coupling to atomic vapors for frequency metrology[28], atom interferometry[29], and free space sensing applications.

We consider wave propagation along a single direction in a uniform medium with linear dispersion, such as a guided wave in a linear waveguide or a slab, or a plane wave in a volume of material. A periodic modulation of the material or waveguide properties along the direction of propagation can linearly couple forward and backward traveling waves whenever phase matching occurs. In the simplest case, modulation with a period $\Lambda_g$ near half the wavelength in the material $\lambda = \lambda_0/n_{eff}$, leads to a direct and, often, strong Bragg reflection of the wave.

However, here we study the effect of the modulation with a period approximately equal to the wavelength. The second-order diffraction, if it is non-zero, leads to the same direct coupling between the forward and backward waves. Meanwhile, if an IM with a zero wavevector is present in the system, the fundamental component of the modulation can also lead to forward-backward coupling occurring through the IM (Fig. 1a). Physically, such an IM can be a combination of waves traveling normal to the original direction of propagation or may originate from local standing wave resonances, and is, generally, lossy. In our experimental photonic grating waveguide (Fig. 1b) with the fundamental periodic modulation wavenumber $k_g=2\pi/\Lambda_g$, the forward and backward traveling slab modes are coupled to each other both through an IM (i.e., surface-normal free-space radiation mode) and directly (due to the first harmonic of the grating modulation with a wavenumber $2k_g$).

We use coupled mode theory [30,31] to describe wave propagation under these general conditions for any such periodically modulated media and to reveal the relation between EPs in a band diagram (BD) and the arising linear wave penetration accompanied by the uniformly distributed energy losses. The CMT equations for the amplitudes $A(z)$ and $B(z)$ of the two modes $A(z)exp(i\omega_B/v_g \cdot z)$ and $B(z)exp(-i\omega_B/v_g \cdot z)$ propagating in opposite directions along the z-axis can be written in the form:

$$\frac{d}{dz}\begin{bmatrix}A\\B\end{bmatrix} = \begin{bmatrix} i\frac{\Delta\omega}{v_g} - \frac{\alpha}{2} & ih_2 \\ -ih_2 & -i\frac{\Delta\omega}{v_g} + \frac{\alpha}{2} \end{bmatrix}\begin{bmatrix}A\\B\end{bmatrix} + h_1(A+B)\begin{bmatrix}-1\\1\end{bmatrix}. \qquad (1)$$

Here $v_g$ is the group velocity of the medium, and $\Delta\omega$ is the angular frequency detuning from the Bragg center frequency $\omega_B = \frac{2\pi v_g}{\Lambda_g}$ defined by the grating period $\Lambda_g$. The wave propagation is described by the linear-dispersion diagonal elements $\pm i\frac{\Delta\omega}{v_g}$ and, generally, the propagation gain or loss coefficient $\alpha$. Henceforth we specifically consider the systems where the propagation loss $\alpha$ is negligible. An off-diagonal coefficient $h_2$ describes the direct coupling of the modes. If the periodic modulation is mirror-symmetric (such as the grating in Fig. 1c), $h_2$ is rendered real-valued by choosing the origin $z = 0$ to coincide with the symmetry plane[31].

The single remaining complex-valued coupling coefficient $h_1$ in Eq. (1) is sufficient to fully describe the coupling effects via an IM when the mode is also mirror-symmetric around $z = 0$. The symmetric IM is coupled with equal phase and amplitude to both propagating modes, i.e., to $A + B$ only. The second term in Eq. (1) fully expresses the change in the dispersion and the losses for each propagating mode due to the IM, as well as the indirect coupling between the modes. The loss through the leaky IM is described by $Re(h_1)$, such that the combined local power loss density is given by $p(z) = 2Re(h_1) \cdot |A(z) + B(z)|^2$.

Coefficients $h_1$ and $h_2$ are generally independent. For example, for a square grating, the first harmonic of the modulation vanishes at the 50 % duty cycle (DC), and for such gratings, $h_2$ can be made arbitrarily small, including $h_2 \lesssim |h_1|$.

For an infinite periodic medium, the well-known Floquet-periodic solutions, i.e., Bloch waves, with complex eigenfrequencies $\Delta\omega_{1,2}$ directly follow from Eq. (1). The corresponding dispersion relationships for our modulated waveguide are illustrated in band diagrams in Fig. 1d-f, where the shading width depicts the imaginary component of the eigenfrequency. For a specific condition $Re(h_2 + ih_1) = 0$, the bandgap at $k = 0$ vanishes, and the BD is described by [24]

$$\frac{\Delta\omega_{1,2}}{v_g} = -ih_1 \pm i\sqrt{Re(h_1)^2 - k^2}, \qquad (2)$$

containing two EPs at $k_{EP} = \pm Re(h_1)$, where eigenvalues and eigenmodes coalesce into one[24]. The group velocity of both modes remains zero within the range of $k$-vectors $|k| < k_{EP}$, where the two modes have equal real eigenfrequency components but significantly differ in the imaginary components describing the losses. The red curves in Fig. 1d-f correspond to a lossy (radiative) mode and the black curves at $k \rightarrow 0$, where imaginary parts are near zero, correspond to the bound state in the continuum (BIC)[32-34]. Here, the radiative mode is the symmetric combination of the propagating modes with equal amplitudes $A = B$, allowing coupling to the symmetric IM (e.g., the free-space radiation). Meanwhile the nonradiative BIC mode is given by an antisymmetric combination with $A = -B$, and the IM coupling is forbidden by symmetry.

Using finite element modeling (FEM) at the optical frequency corresponding to $\lambda_0 = 780$ nm vacuum wavelength, we find that in a 250 nm thick silicon nitride (SiN$_x$) photonic slab waveguide modulated by a shallow uniform grating etched 85 nm vertically into it (Fig. 1c), and cladded by SiO$_2$, the EP condition $Re(h_2 + ih_1) = 0$ is realized for the groove DC = DC$_{EP}$ ≈ 0.527 (Fig. 1e, Fig. S2). Band diagrams simulated using FEM (scatter plot markers in Fig. 1e) strongly agree with the results obtained from CMT (solid curves). Once the DC is significantly detuned from the DC$_{EP}$ the conventional bandgap opens, removing the degeneracy (Fig. 1d, f).

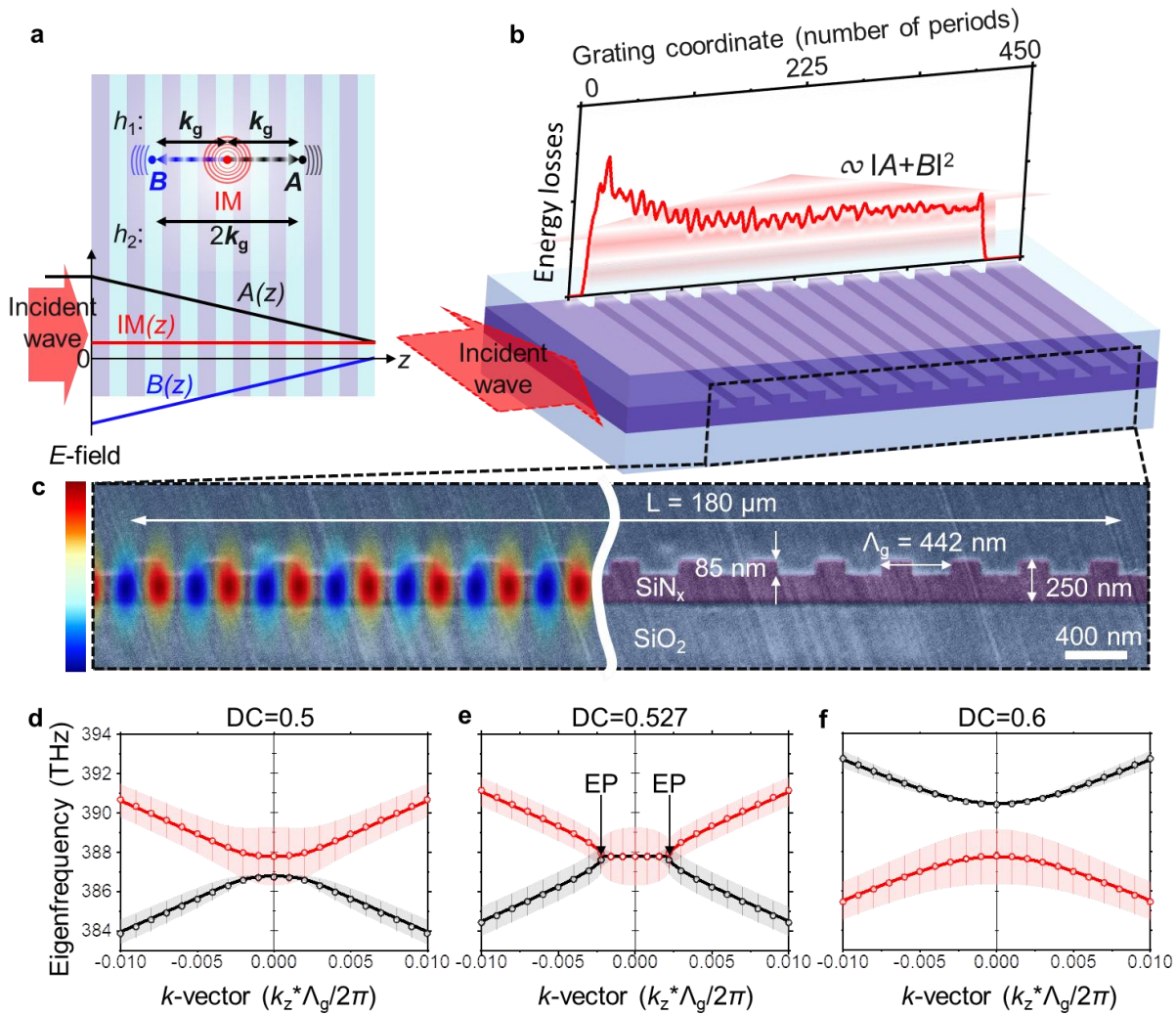

Figure 1. Linear wave penetration and uniform energy losses across a periodically structured medium with exceptional-point-containing band diagrams. (**a**) The forward, *A*, and backward, *B*, traveling waves are coupled via a stationary Intermediary Mode, IM, by the modulation reciprocal vector $k_g$. The first harmonic at $2k_g$ couples *A* and *B* directly. At the exceptional point (EP), the amplitudes *A* and *B* are linear, and IM is constant across the full width of the medium. (**b**) An ≈ 50 % duty cycle (DC) photonic grating waveguide excited from one side exhibits uniform energy dissipation, projecting an experimentally observed beam of uniform intensity. (**c**) A false-color cross-

sectional scanning electron microscopy (SEM) image showing the transverse E-field component (normal to the image plane) simulated using FEM and a waveguide structure comprising a 250 nm thick SiN$_x$ slab with 85 nm deep grooves, nominal period $\Lambda_g$ = 442 nm, and the length is $L \approx 180$ µm (400 periods). (**d-f**) Band diagrams for different duty cycles DC = 0.5, 0.527, and 0.6, simulated using FEM (scatter plot) and calculated with CMT (solid curves). The solid curves and scatter plot markers depict the real parts of eigenfrequencies, and the vertical dimensions of shaded regions and vertical lines show the imaginary eigenfrequencies.

We now consider how the waves penetrate into a macroscopic layer of such medium having a large but *finite* length *L*, experimentally implemented by a uniform photonic grating with a large but fixed number of periods, illuminated by a guided photonic slab mode incident from one side. We solve the CMT Eq. (1) for continuous wave (CW) illumination, setting the boundary condition as $A(z = 0) = 1$ and $B(z = L) = 0$. For CW illumination the BIC frequency, given by $\frac{\Delta\omega}{v_g} = h_2$ corresponds to an EP. (See Supplementary Information Section 1). For frequencies far away from the $\frac{\Delta\omega}{v_g} = h_2$ condition, both *A* and *B* decay exponentially from *z* = 0. However, as $\frac{\Delta\omega}{v_g}$ approaches $h_2$, light penetrates much deeper into the medium, and both *A* and *B* become simple linear functions of *z* at $\frac{\Delta\omega}{v_g} = h_2$, and the power lost into the IM becomes constant and independent of *z*. The power is maximized for a 52.7 % DC grating with EPs in the band diagram, i.e., $Re(h_2 + ih_1) = 0$, and the amplitudes of two modes and the radiated power are given by:

$$A(z) = \frac{1 - Re(h_1)(z-L)}{1 + Re(h_1)L}, \tag{3a}$$

$$B(z) = \frac{Re(h_1)(z-L)}{1 + Re(h_1)L}, \tag{3b}$$

$$p(z) = 2Re(h_1) \cdot |A(z) + B(z)|^2 = \frac{2Re(h_1)}{(1+Re(h_1)L)^2}, \quad (3c)$$

i.e., an incident wave undergoes deep, linear penetration accompanied by the energy loss that is *uniformly distributed across arbitrary finite thickness L* of a homogeneous periodically structured medium at the frequency tuned to the EP.

In agreement with the CMT, the FEM and the finite-difference time-domain method (FDTD) simulations for the *finite length* waveguide with EPs confirm the constant intensity profile of a radiated free-space beam occurring near the EP frequency (Fig. 2, also cf. Fig. S3). For frequency-detuned light (e.g., $\lambda_0$ = 778 nm) incident on the DC ≈ 50 % grating, the radiated beam has a typical exponential decay profile (Fig. 2a, top plot) associated with the constant decay rate of the guided mode $A(z)$ as the light couples out into free space. Conversely, tuning the frequency to the EP ($\lambda_0$ = 773 nm) leads to a constant-amplitude radiation profile over the full length of the modulated section (Fig. 2a, bottom plot). Notwithstanding the small initial intensity spike associated with scattering due to the mode and effective index mismatch at the slab-grating interface, the radiated power shows no decrease until the far end of the grating, over a distance several times longer than the length scale of the conventional exponential decay in the detuned case. The red arrows in Fig. 2a insets depict the free-space radiation direction.

The qualitative difference of the new wave propagation regime becomes particularly clear when considering penetration into the EP-containing structured medium of different lengths. Figure 2b compares the intensity profiles radiated by the grating waveguides of different lengths, contrasting the conventional regime at an incident guided wave frequency detuned from the EP with the deep linear penetration regime at the EP frequency. In stark contrast with the conventional exponential decays, which are independent of the waveguide length (top panel), the uniform power densities radiated into free-space at the EP frequency (bottom panel) vary depending on the full length $L$ of the grating waveguide, as predicted by Eq. (3c). In all cases, the modulated section of the waveguide is excited by the *same* power incident from one side ($z < 0$) only.

Qualitatively, the deep penetration into finite layers is intimately related to the lossless, spectrally narrow BIC mode in the infinite medium. Indeed, the wavelength dependence of the radiated power near the structured waveguide output end, opposite to the incident wave (Fig. 2c,d), reveals the narrow-band character of the constant-intensity radiation, with the spectral bandwidth decreasing with $L$ (approximately $\propto L^{-2}$) (inset of Fig. 2d) in agreement with CMT (see Supplementary Information for details). Hence for larger lengths of periodic materials, the device operation becomes extremely sensitive to small parameter variations. As the frequency transits the EP frequency, the outcoupled angle of the free-space beam undergoes fast variation around the surface-normal, flipping sign, clearly noticeable in Extended Data Fig. 1a as a double kink. This angle variation corresponds to the frequency sweep through the BD of the infinite-thickness medium from the top photonic band to the lower photonic band across both EPs (Extended Data Fig. 1b).

The detailed analysis of normalized free-space amplitudes as a function of wavelength and position $z$ for gratings with DC = 0.5, 0.527, and 0.55 using CMT and FDTD (Fig. S3) implies the deep wave penetration is observed in each grating. However, only the grating tuned to operate at EP frequency (DC = 0.527, middle panel in Fig. S3) manifests the linear energy losses in contrast to other counterparts that show distorted radiation profiles (top and bottom panels in Fig. S3), highlighting the importance of the EP for linear wave penetration (Extended Data Fig. 2).

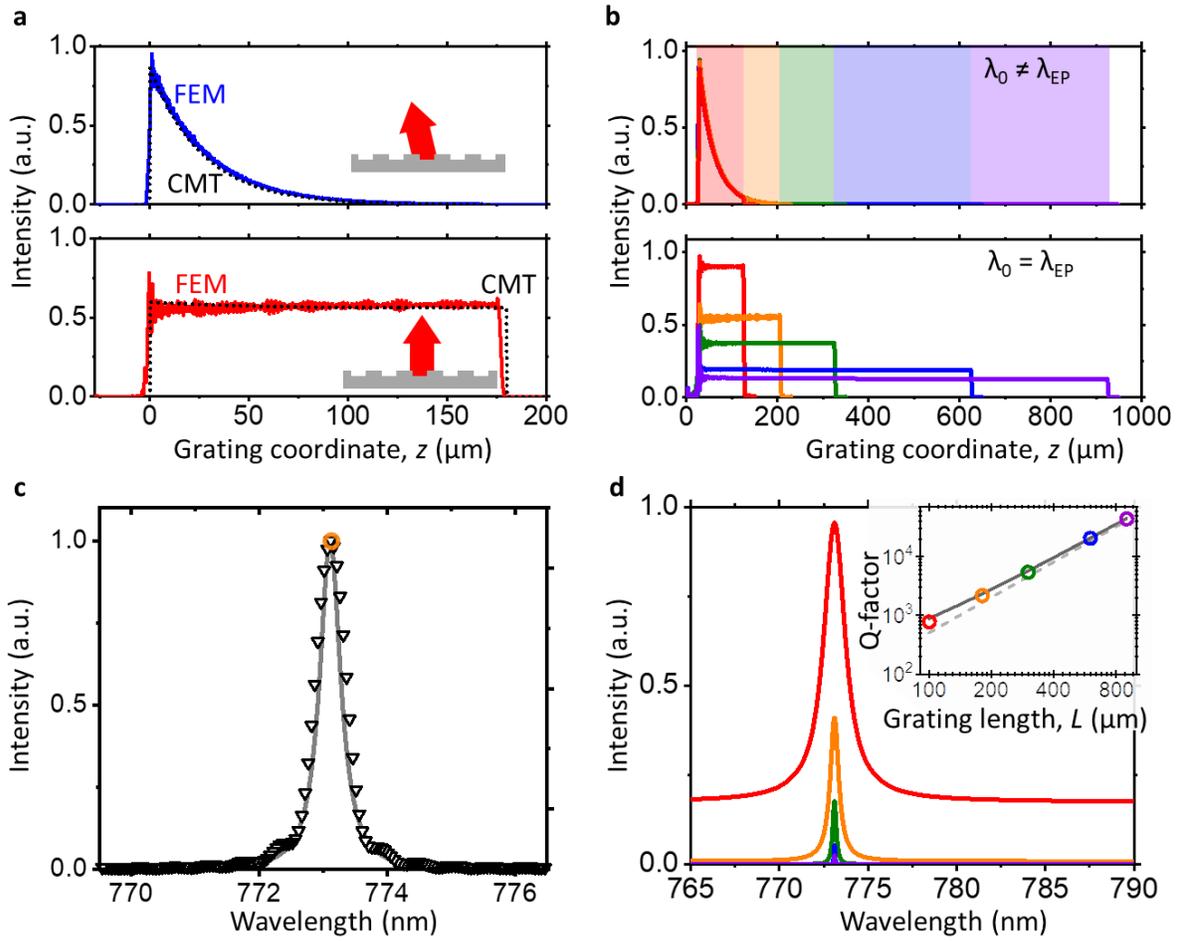

Figure 2. CMT and numerical analysis of the light radiated by a periodically structured SiN$_x$ slab waveguide. (**a**) The radiated intensity dependence on the location $z$ along the waveguide, at $\lambda_0$ = 778 nm (detuned from EP condition, top curve) and $\lambda_0$ = 773 nm (tuned into EP, $\lambda_{EP}$ = n$_{eff}$Λ$_g$, bottom curve). Modulated section is 0 μm < $z$ < 180 μm. The insets illustrate the direction of the free-space beam. (**b**) FEM-simulated radiated intensity for grating lengths $L$ = 100 μm (red curve), 180 μm (orange curve), 300 μm (green curve), 600 μm (blue curve), and 900 μm (violet curve). The top plot depicts for comparison the conventional decaying intensity profiles for the wavelength $\lambda_0$ = 778 nm ($\lambda_0 \neq \lambda_{EP}$), where the radiated intensity decay is independent of length, and hence all curves lie on top of each other. In contrast, the bottom plot shows decay-free profiles at the wavelength $\lambda_0$ = 773 nm ($\lambda_0 = \lambda_{EP}$),

where the radiated intensity depends on the grating length. (**c**) FDTD-simulated intensity spectrum (scatter plot) and CMT results (solid curve) near the grating end ($z$ = 170 µm, $L$ = 180 µm) for EP duty cycle DC = 0.527. Little spectral broadening at the peak bottom is attributed to the finite pulse duration in FDTD simulations. An orange circle marks the linear energy wave penetration. (**d**) Intensity spectra calculated using FEM at the grating far end, $z = L$, opposite to the incident wave for different lengths $L$ = 100 µm, 180 µm, 300 µm, 600 µm, and 900 µm. Log-log inset shows the effective quality factor $Q = \lambda_0$/FWHM increasing with $L$. The solid line is the CMT model, and the dotted line depicts the $Q \propto L^2$.

To verify experimentally the deep penetration and uniformly distributed energy loss in periodic media, we microfabricate uniform grating waveguides with the dimensions specified in Fig. 1 using the common nanofabrication methods (See Methods). The test gratings range in lengths from the 45 µm to 250 µm with nominal DC = 0.4, DC = 0.5, DC = 0.53, and DC = 0.6. To excite gratings from one end, we launch a collimated TE polarized slab mode using an evanescent coupler with a spatially apodised gap formed between an input waveguide and the slab [26] (Fig. 3a). The slab mode is incident orthogonally to the grating lines. Images of the free-space radiation at input wavelengths away from the EP frequency (Fig. 3b) and tuned to the EP frequency (Fig. 3c) taken from the $L$ = 180 µm grating waveguide with DC carefully tuned near 0.5 reveal the exponentially decaying and constant-intensity profiles, respectively. In particular, it is clear from Fig. 3c that the light outcouples uniformly across the entire grating area for the wavelength tuned to the EP. Conversely, for the detuned case, all measurable optical power outcouples at the front portion of the grating (Fig. 3b). The corresponding integrated plots of radiated intensity and its wavelength dependence at the end of the grating ($z \approx 170$ µm) are depicted in Figs. 3b-d and agree with the corresponding FEM and FDTD results (Figs. 2a and 2c). The experimental Q factor is 2570 ± 110, which is close to the value predicted by the CMT. The value and the uncertainty are obtained from the Lorentzian fit to the experimental spectrum in Fig. 3d. Top-hat intensity beams surface-normally projected by 45 μm, 100 μm, 180 μm, and 250 μm long gratings with DC ≈ 0.5 (Extended Data Fig. 3) fall in with simulations as well (Fig. 2), experimentally revealing the novel linear wave penetration regime in the finite thickness non-Hermitian wave media with EPs.

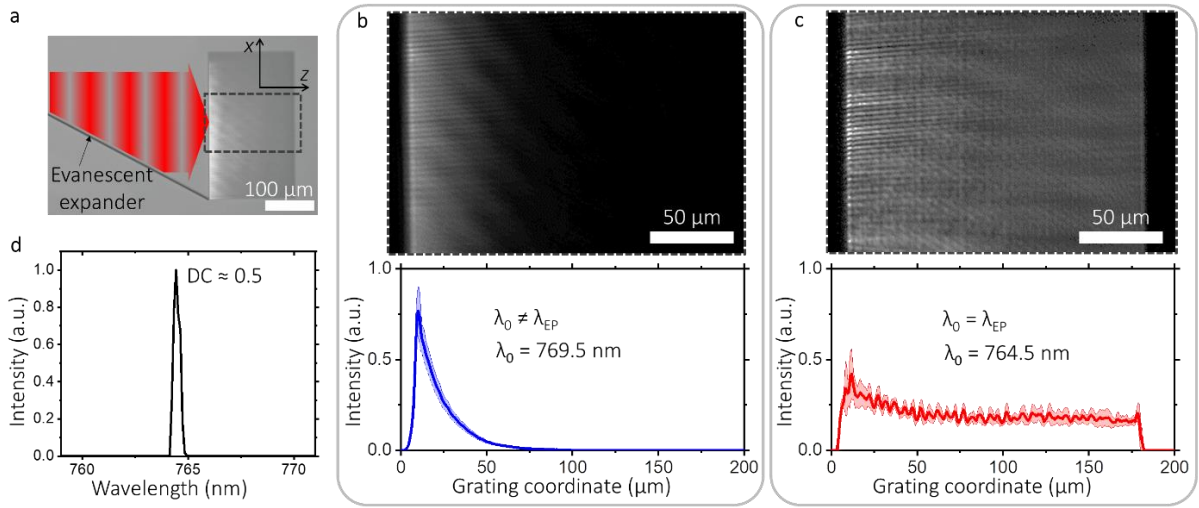

Figure 3. Experimental observation of the deep linear light wave penetration with constant-intensity radiation profiles. (**a**) A top-view optical microscope image of the 180 μm long grating waveguide with an incident collimated slab mode (schematically shown in red) launched into it by an evanescent expander. The grating dimensions along the *z*-axis are the same as in Fig. 1c. Optical images of the beams projected from the same device and the corresponding averaged intensity profiles at the wavelength detuned from (**b**) and tuned into (**c**) the EP frequency. The curves are obtained from the corresponding images by integrating intensity along the *x*-axis. One standard deviation statistical uncertainty shown by the shaded areas is obtained by analyzing multiple separate slices of the image. (**d**) Spectral dependence of free-space radiation at the end of the grating ($z \approx 170$ μm) with DC ≈ 0.5.

We observe good agreement between the experiment, CMT, and numerical simulation results across a range of frequencies (Fig. 4, Fig. S3). Mapping the normalized free-space amplitude, as a function of the wavelength and the coordinate $z$, using CMT (Fig. 4a), FDTD (Fig. S3b), and experimental data (Fig. 4b), verifies the decay-free profile at the EP-tuned frequency and an exponentially decaying intensity upon the wavelength detuned from that condition. The maps for the grating with EP, i.e., DC = $DC_{EP}$, show approximately equally quick exponential decays in intensity vs. the coordinate $z$, upon red and blue wavelength detuning from the EP (middle panels in Fig. 4 and Fig. S4). We observe red-blue asymmetric exponential decays when DC is detuned up or down from $DC_{EP}$, such as in top and bottom rows in Fig. 4 and Fig. S4. These asymmetric spectral plots with DCs detuned from EPs in corresponding BDs, i.e., DC ≈ 0.4 and DC ≈ 0.6 in Fig. 4, reveal Fano-like peaks. Qualitatively, these complex radiation patterns with Fano-like asymmetry result from the spectral overlap, coupling and energy transfer between the spatially nonuniform high-Q BIC-like and the low-Q leaky mode centered at slightly different wavelengths (Figs. 1d and 1f) in the finite system where they are no longer the eigenmodes.

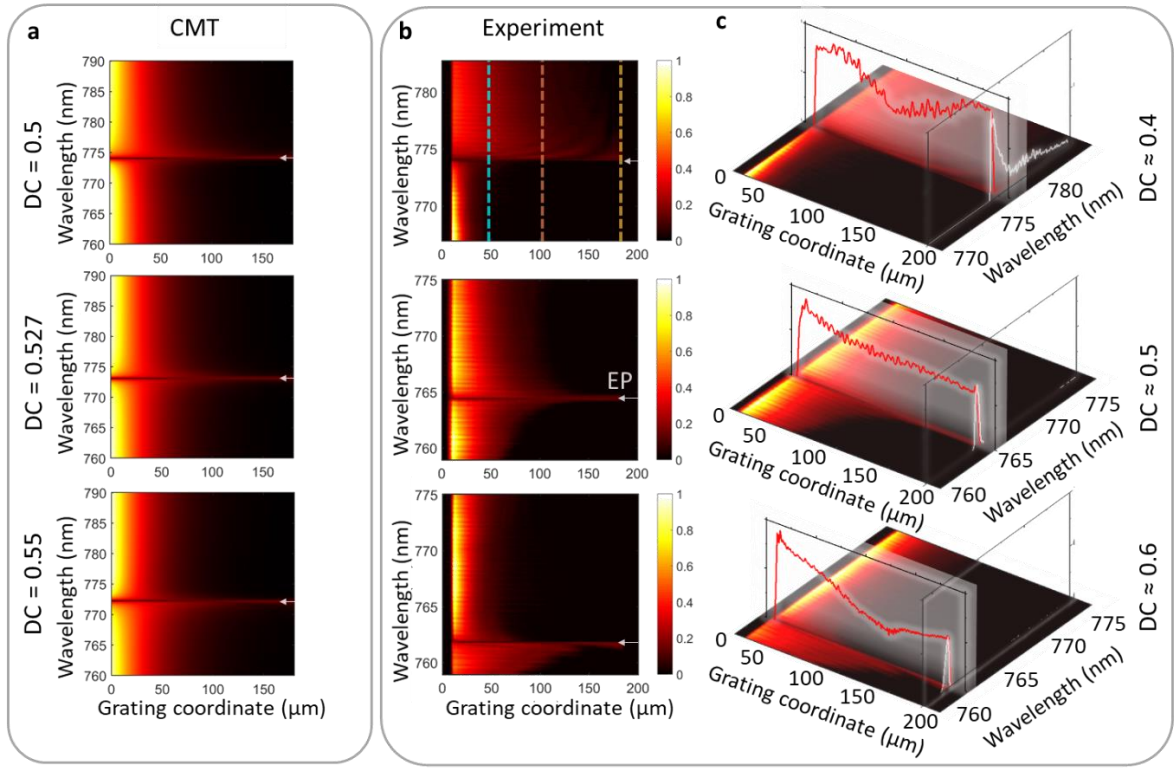

Figure 4. Radiation loss profiles in non-Hermitian periodically structured media. (**a**) The normalized amplitudes radiated into free space by a single-side excited grating waveguide as a function of the wavelength and position $z$, for different grating duty cycles (DC = 0.5, 0.527 and 0.55, top to bottom) obtained using CMT. (**b**) Experimentally measured radiated amplitudes collected from gratings with DC ≈ 0.4, 0.5, and 0.6 (top to bottom). White arrows point out the deep wave penetration. Blue, red, and yellow dashed lines show the coordinates where spectral profiles shown in Fig. S4 are collected. (**c**) Experimental spectra (white curves) at the grating ends ($z ≈ 180$ μm) and amplitude profiles (red curves) drawn through the maxima of the corresponding spectra for different DC gratings. Experimentally the linear penetration is observed for a grating with the nominal DC = 0.5 vs. the CMT-predicted DC = 0.527, corresponding to ≈ 12 nm difference in the grating groove width, which is within the fabrication variation.

From a practical perspective, proper tuning into the linear wave penetration regime comprises three necessary conditions: (i) the uniformity of the periodic structure across the device, (ii) the presence of the EP in a band diagram, and (iii) the optical frequency tuned to the EP. For 180 μm long gratings manifesting EP in a band diagram, the bandwidth of linear wave penetration is $\Delta\lambda \approx 0.1$ nm (Figs. S4 – S6). For small DC detuning away from the EP condition, the deep penetration is still evident at the frequency approximately corresponding to the BIC. However, the amplitudes and the dissipated power vary spatially and are no longer linear and constant (Fig. 4c, top and bottom panels, Figs. S7, S8) in stark contrast to linear wave penetration (Fig. 4c, middle panel) observed in the gratings with EP.

Finally, periodic medium uniformity should be maintained with grating line widths within a few-nanometer range across the device lengths to ensure a linear wave loss propagation. Proximity effects in electron beam lithography patterning of the gratings may naturally result in the variation of the exposure and the corresponding reduction in the line width near the grating edges. To obtain uniform profiles, proximity effects have been carefully corrected in our exposures. Suboptimal exposures resulted in DC nonuniformities (Fig. S10 a) that can be approximated by the distribution shown in Fig. S10 b. The resulting free-space beam profiles manifest a bright spike at the leading edge of the grating quickly decaying to a constant intensity across the grating in qualitative agreement with simulations for the nonuniform gratings (Fig. S10 c, d). This effect also explains small distortions from ideal decay-free profiles seen in the experimental data in Fig. 3 c and Extended Data Fig. 3.

To conclude, we have discovered theoretically and verified experimentally a novel wave penetration regime into finite layers of uniform, periodically structured non-Hermitian wave media with EPs, characterized by arbitrary deep wave penetration with a linear spatial dependence of the wave amplitudes and a uniformly distributed energy loss. This regime arises when forward and backward traveling waves are coupled by the periodic modulation of the medium index directly and indirectly through a third, stationary lossy intermediate mode. It is experimentally observed in a photonic slab waveguide periodically structured by a shallow etched grating with the duty cycle of ≈ 50 %, and a period that couples the guided waves to a surface-normal free-space wave, resulting in radiation loss. The uniformly distributed radiation intensity into free space was observed for several grating lengths > 400 wavelengths long. The spectral bandwidth of the unusual penetration regime is shown to decrease proportional to the square of the medium thickness (waveguide length). Our findings advance fundamental understanding of wave propagation in non-Hermitian dispersion-engineered wave media across physical domains.

METHODS

**Finite element modeling simulations.**

*Band diagram analysis*

The band diagram analysis is conducted using a FEM approach. We solve the eigenvalue problem for a 2D unit cell where Floquet periodic conditions are defined on opposite cell boundaries. The unit cell comprises a 250 nm thick $SiN_x$ slab with top and bottom claddings made of $SiO_2$. The top cladding is backed with a perfectly matched layer. For modeling, we assume the lossless materials with refractive indices of 2.01 ($SiN_x$) and 1.45 ($SiO_2$) around the 773 nm vacuum wavelength. The grating depth and the period are 85 nm and $\Lambda_g$ = 442 nm, respectively.

*Finite grating simulations*

To perform the FEM simulations, we define the 2D finite grating structure and 3 ports: 2 ports for photonic slab modes at the opposite grating ends and one free space port above the grating. We launch the propagating slab mode $A(z)$ from one side of the 180 μm long grating at $z$ = 0 (Fig. 1a) and measure the electric field profile 3 $\mu$m above the grating surface in free space. Then the complex free-space electric field is convolved with a 3.2 μm wide focal spot corresponding to a 0.3 NA objective used in the experiment.

**Sample fabrication.**

The standard nanofabrication techniques are employed to fabricate all photonic grating samples, as shown in the flowchart (Extended Data Fig. 4). Continuous films of ≈ 2.9 μm thick $SiO_2$ and nominally 250 nm $SiN_x$ are grown on a Si wafer using thermal oxidation and low-pressure chemical vapor deposition (LPCVD), respectively. The electron-beam lithography followed by the reactive-ion etching is implemented two times sequentially to form the pattern of ≈ 85 nm deep partially etched gratings with different DCs in the SiN$_x$ slab and fully etched SiN$_x$ waveguides and evanescent expanders. ≈ 3 µm thick SiO$_2$ cladding is formed on top of the written patterns employing LPCVD. Top-view and cross-sectional scanning electron microscopy as well as atomic force microscopy are implemented to routinely characterized grating dimensions and photonic waveguides.

**Experimental set-up.**

We use a tunable fiber-coupled laser to feed a TE$_0$ single-mode on-chip waveguide via an inverted-taper coupler. To obtain a wide collimated photonic slab mode, we expand the TE$_0$ waveguide mode using an evanescent coupler with a spatially apodised gap formed between the spatially uniform waveguide and the slab [26] (Fig. 3a). Since the gap size determines the coupling strength between both modes, the slab wave with arbitrary-defined intensity distribution can be engineered by apodizing the gap along the expansion region. We vary the gap to obtain a slab mode with a 200 µm full width at half maximum (FWHM) Gaussian intensity profile. The collimated photonic slab mode is launched at the angle relative to the waveguide determined by the phase matching relation between the waveguide and slab modes and is normally incident onto the grating from one side only. We capture the images of the free-space beam emanating from the grating using a microscope equipped with an NA = 0.3 objective (Extended Data Fig. 5), which is mounted above the chip. The grating projects the slab mode into free-space at the near-normal angle, defined by the phase matching condition $\beta_0 - \frac{2\pi}{\Lambda_g}m = \frac{2\pi}{\lambda}\sin\theta$, with $m = 1$. The angle is experimentally quantified by defocusing the microscope a few millimeters above the chip surface (Fig. S11).

ACKNOWLEDGEMENTS

Dr. Alexander Yulaev acknowledges support under the Professional Research Experience Program (PREP), funded by the National Institute of Standards and Technology and administered through the Department of Chemistry and Biochemistry, University of Maryland. Authors also thank Dr. Thomas LeBrun, Dr. Henri Lezec, Dr. Jin Liu, Dr. J. Alexander Liddle, and Dr. Erik Secula for reading the manuscript and making insightful comments.

AUTHOR CONTRIBUTIONS

AY and SK contributed to the project equally. SK conceived the idea. AY and SK performed FEM and FDTD simulations. DAW and AY fabricated samples. QL, BJR and KS contributed to the experimental device design and fabrication process development. AY and SK characterized samples. AY and VAA developed theoretical analysis. AY, SK, and VAA contributed to the data interpretation. VAA supervised the project. The manuscript was written through the extensive contributions of all authors. All authors have approved the final version of the manuscript.

COMPETING INTERESTS

The authors declare no competing financial interests

ADDITIONAL INFORMATION

Supplementary Information is available for this paper.

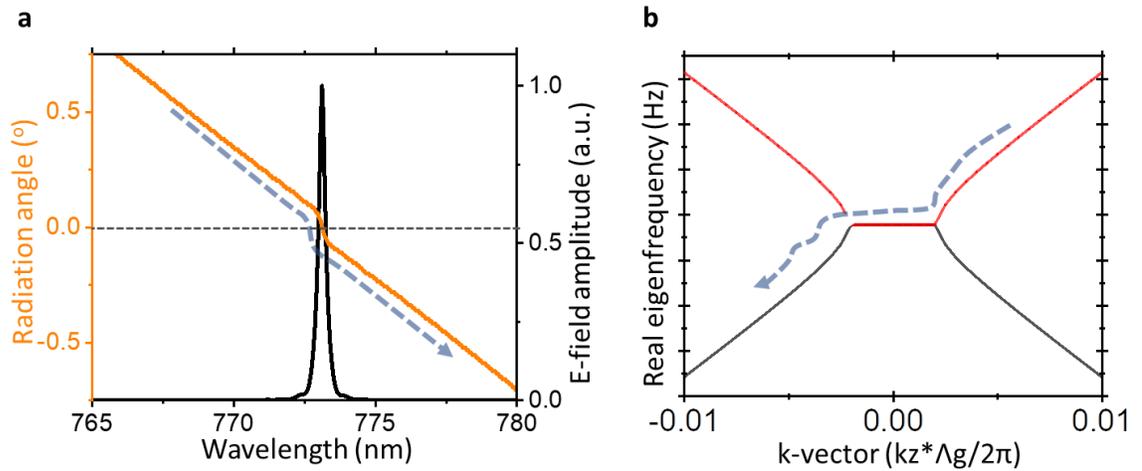

Extended Data Fig. 1. Free-space beam angle and intensity dependences on the optical wavelength. (**a**) The orange curve shows the radiation angle as a function of the wavelength for DC = 0.527. The black dashed line indicates 0°. (**b**) The corresponding BD for infinite grating. The dashed blue arrows depict the wavelength tuning. Once the wavelength crosses the EPs in the BD (panel b), the outcoupling angle experiences high variation about the chip's normal, that is noticeable as a double kink in the wavelength dependence of the outcoupled angle (panel a).

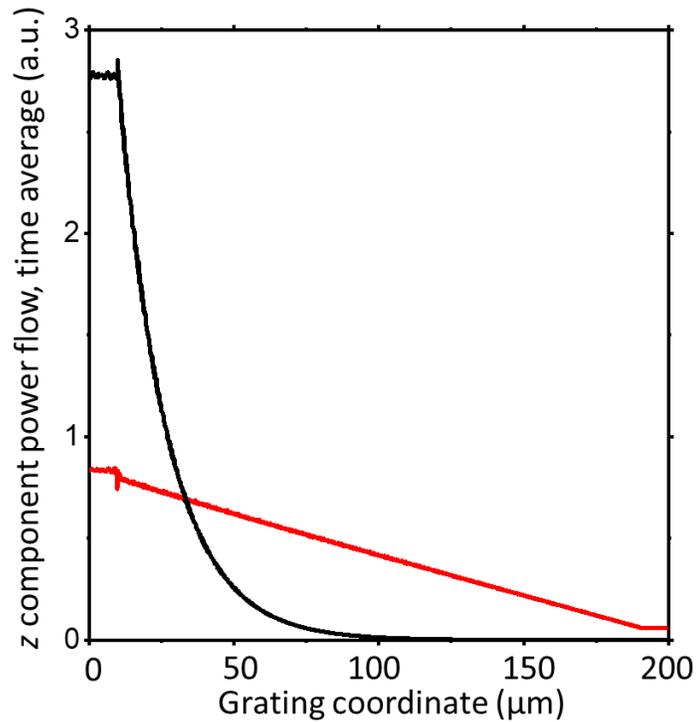

Extended Data Fig. 2. Power flow simulation indicating spatially uniform energy losses in non-Hermitian periodically structured media excited from one side. FEM simulated time-averaged power flow along the uniform periodically structured waveguide in the positive $z$-direction vs. the coordinate $z$ for the optical frequency tuned to the EP wavelength (red curve) and detuned (black curve) from the EP wavelength. The waveguide is excited by a wave incident from the left side at $z = 10$ μm.

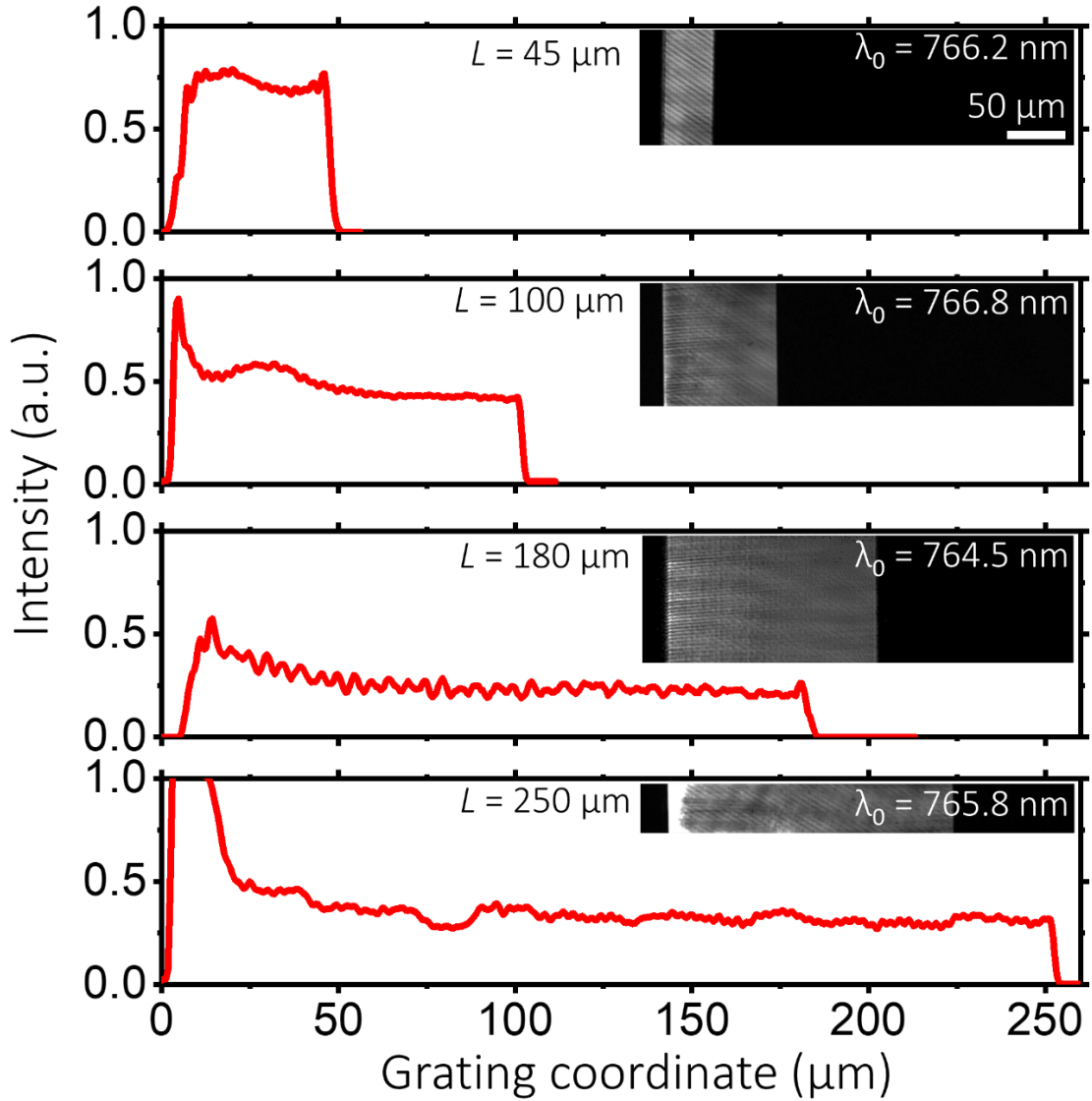

Extended Data Fig. 3. Linear radiative energy losses experimentally observed in 45 μm, 100 μm, 180 μm, and 250 μm long gratings with ≈ 50 % DC. Insets depict top-view optical images of the projected free-space beams. The saturated ≈ 20 μm wide spike at the grating input is due to the fabrication imperfections related to the electron lithography proximity effect (Fig. S10).

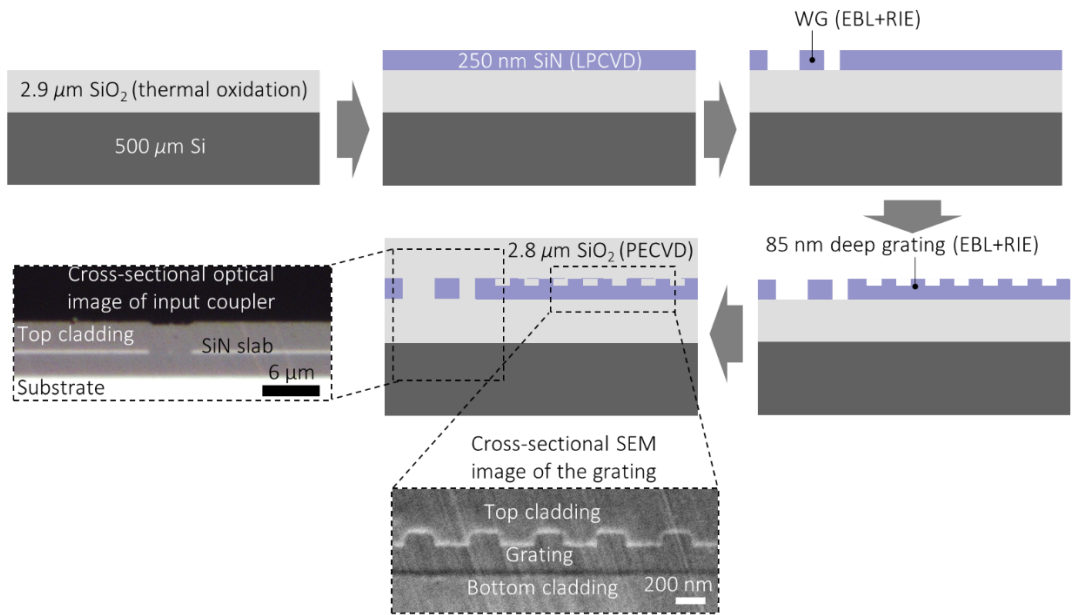

Extended Data Fig. 4. Sample fabrication flowchart.

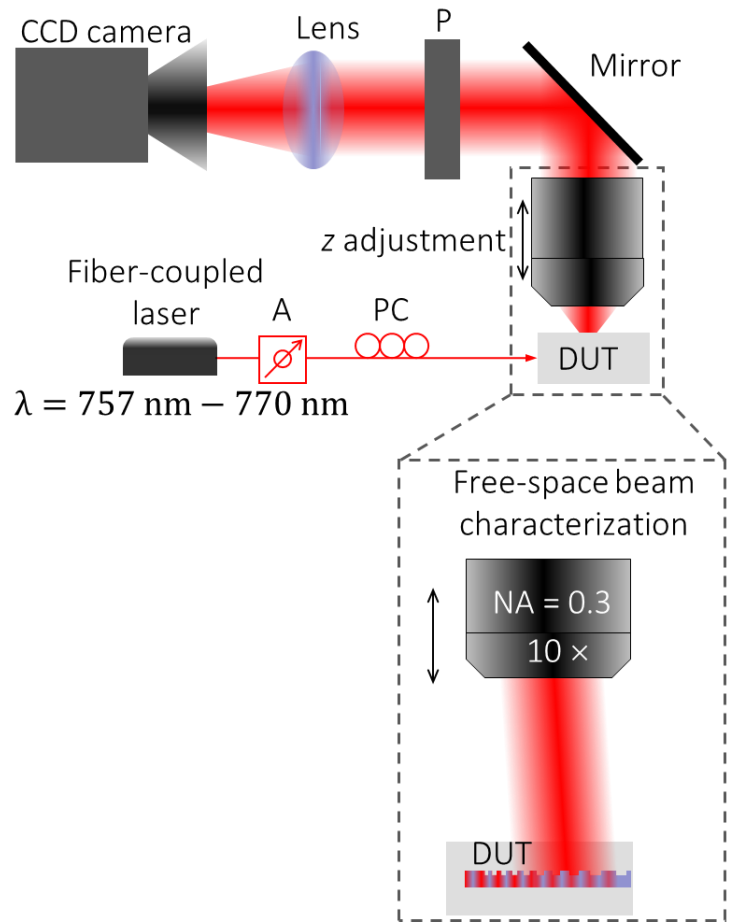

Extended Data Fig. 5. Experimental apparatus. A – fiber attenuator, PC – fiber polarization controller, DUT – device under test, P – free-space polarizer.

Supplementary information for

# Exceptional points in lossy media enable decay-free wave propagation

*Alexander Yulaev[1,2,†], Sangsik Kim[1,3,†], Qing Li[1,4], Daron A. Westly[1], Brian J. Roxworthy[1,5],*

*Kartik Srinivasan[1], and Vladimir A. Aksyuk[1]\**

[1]*Physical Measurement Laboratory, National Institute of Standards and Technology,*

*Gaithersburg, MD 20899, USA.*

[2]*Department of Chemistry and Biochemistry, University of Maryland, College Park, MD 20742,*

*USA.*

[2]*Department of Electrical and Computer Engineering, Texas Tech University, Lubbock, TX*

*79409, USA.*

[4]*Department of Electrical and Computer Engineering, Carnegie Mellon University, Pittsburgh,*

*PA 15213, USA.*

[5]*Aeva, Inc. 555 Ellis Street, Mountain View, CA 94043, USA.*

[†]*These authors contributed equally to this work*

*\*Corresponding author: vladimir.aksyuk@nist.gov*

1. **General CMT equation solution and transition to constant-intensity profile**

We start from CMT equations describing the coupling between two counterpropagating modes $A(z)exp(i\omega_B/v_g \cdot z)$ and $B(z)exp(-i\omega_B/v_g \cdot z)$ with slowly varying amplitudes $A(z)$, $B(z)$ across the periodically structured medium, and $E_0(z)$, an intermediary mode amplitude. The intermediary mode is stationary, having a near-zero wavevector projection on the $z$-axis. Here we generalize the specific approach developed in Refs. [1,2] for grating waveguides coupled to free space. Replacing the free space radiation mode by a general lossy intermediary mode at near-zero wavevector is the key concept enabling generalization to a wide range of systems supporting propagating waves.

For the purposes of this work we impose two additional constraints on the system. The first one is that the modulation of the wave medium is mirror symmetric. This allows picking the origin $z = 0$ at the symmetry plane such that the direct coupling coefficient $h_2$ is real-valued. Secondly, we consider only such intermediary modes that have the same mirror symmetry. Being stationary, the intermediate mode can in general be coupled to both of the stationary combinations of the two traveling waves, characterized by amplitudes $A + B$ and $A - B$. However, a symmetric intermediary mode couples only to the symmetric combination. The resulting coupled mode equations take the following simple form

$$E_0(z) = \sqrt{h_1}(A + B), \tag{1}$$

$$\frac{d}{dz}\begin{bmatrix}A\\B\end{bmatrix} = \begin{bmatrix} i\frac{\Delta\omega}{v_g} - \frac{\alpha}{2} & ih_2 \\ -ih_2 & -i\frac{\Delta\omega}{v_g} + \frac{\alpha}{2} \end{bmatrix}\begin{bmatrix}A\\B\end{bmatrix} + \sqrt{h_1}E_0(z)\begin{bmatrix}-1\\1\end{bmatrix}. \tag{2}$$

Here, $h_2$ and $\alpha$ are the coefficients describing the coupling between the two counter-propagating modes with amplitudes $A(z)$ and $B(z)$ (real-valued), and their propagation loss/gain, respectively. $\sqrt{h_1}$ is the complex-valued coefficient describing the coupling of these modes to and from the intermediary mode with amplitude $E_0(z)$ (e.g., the free-space radiation in our experiment), which also accounts for the losses through that mode. $\Delta\omega$ and $v_g$ are the optical angular frequency difference relative to the Bragg frequency at zero index modulation and the group velocity, respectively. The +1 and -1 in the second term account for the different propagation directions of the propagating modes.

Substituting Eq. (1) into Eq. (2), we get

$$\frac{d}{dz}\begin{bmatrix}A\\B\end{bmatrix} = \begin{bmatrix} i\frac{\Delta\omega}{v_g} - \frac{\alpha}{2} & ih_2 \\ -ih_2 & -i\frac{\Delta\omega}{v_g} + \frac{\alpha}{2} \end{bmatrix}\begin{bmatrix}A\\B\end{bmatrix} + h_1(A+B)\begin{bmatrix}-1\\1\end{bmatrix}. \qquad (3)$$

From Eq. (3), the corresponding eigenvalues defining the periodic system band diagram in case of no material loss and gain are given by

$$\frac{\Delta\omega}{v_g} = -ih_1 \pm \sqrt{(h_2 + ih_1)^2 + k^2}, \qquad (4)$$

where $k$ is a wave number. The choice of the sign in Eq. (4) defines either a leaky radiative mode or a bound state in the continuum (BIC) when $A$ and $B$ modes interact to enhance or cancel radiation, respectively.

Eq. (4) defines the complex frequency eigenvalues $\Delta\omega$ for the infinite Floquet periodic problem with real-valued wavevectors $k$. To describe the continuous-wave experiments performed for specific real-valued optical frequencies, it is also useful to consider the eigenmodes of Eq. (3) for given real-valued frequency parameter, with complex-valued wavevectors $k$ as eigenvalues, given by

$$k = \pm\sqrt{(\Delta\beta + ih_1)^2 - (h_2 + ih_1)^2} \qquad (4')$$

with real-valued parameter $\Delta\beta = \frac{\Delta\omega}{v_g}$. These functions are illustrated in Fig. S1 for three different values of $h_2$ such that $Re(h_2 + ih_1)$ is >0, <0 and =0. For $\Delta\beta \neq h_2$ there are two eigenmodes with spatial dependencies $\propto exp(\pm ikz)$, and the solution inside a finite length of the periodic structure is given by their linear combination satisfying the boundary conditions at the structure's ends.

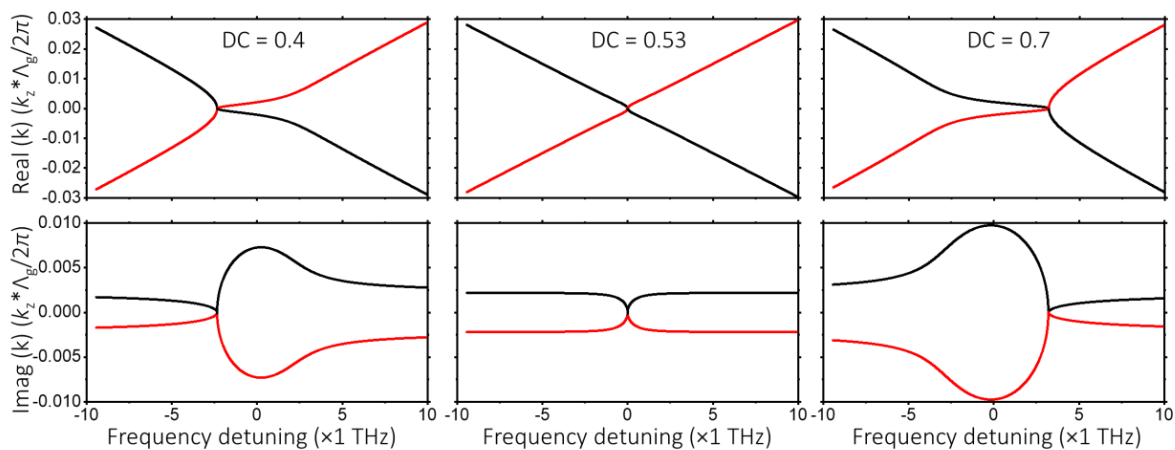

Fig. S1. Calculated dependences of real (top) and imaginary (bottom) components of complex-valued wavevector eigenvalues on the real-valued frequency. The frequency detuning is defined with respect to the Bragg condition ($\Delta\beta = 0$). The values of $h_2$ are chosen to satisfy $Re(h_2 + ih_1)$ is <0, =0 and >0 conditions that are realized at DC = 0.4, 0.53, and 0.7, respectively.

For the finite periodic structure of length $L$ excited from one side, the boundary conditions become $A(z = 0) = 1$ and $B(z = L) = 0$, and the grating mode profiles take the form

$$A(z) = \frac{ik \cosh ik(L-z) - (i\Delta\beta - h_1) \sinh ik(L-z)}{ik \cosh ikL - (i\Delta\beta - h_1) \sinh ikL}, \quad (5a)$$

$$B(z) = \frac{i(h_2 + ih_1) \sinh ik(L-z)}{ik \cosh ikL - (i\Delta\beta - h_1) \sinh ikL}. \quad (5b)$$

However, $\Delta\beta = h_2$, is an exceptional point, where both eigenvalues $k$ become equal to 0, and both eigenmodes coalesce to unity, becoming spatially independent. At this EP, occurring at the bound state in the continuum (BIC) frequency, the solution is no longer a linear combination of two different eigenmodes, but rather Eqs. (5a-5b) simplify to the linear expressions:

$$A(z) = \frac{ih_2 - h_1}{1 - (ih_2 - h_1)L} z + 1, \quad (6a)$$

$$B(z) = \frac{i(h_2 + ih_1)}{1 - (ih_2 - h_1)L} (L - z). \quad (6b)$$

The power loss from the propagating modes through the intermediary mode (e.g. radiated into free space) is given by $p(z) = 2Re(h_1) \times |A(z) + B(z)|^2$, and according to Eqs. (6a-6b), becomes independent of the coordinate $z$, enabling uniform radiation losses across the whole length of $L$. To plot solutions using CMT, we use $h_1 = 31270$ m$^{-1}$ and the following values of $h_2 = -17800$ m$^{-1}$, 0 m$^{-1}$, and 15400 m$^{-1}$ for DC =0.5, DC = 0.527, and DC = 0.55, respectively. All values are chosen based on fitting BDs in Figs. 1 d-f.

To estimate the bandwidth of the decay-free radiation, we consider operation slightly detuned from the constant-intensity regime, i.e., $\Delta\beta = h_2 + \delta\beta$. Hence, we can incorporate the following approximations $k \approx \sqrt{2iRe(h_1)\delta\beta}$ and truncate Taylor expansion of harmonic functions as $\frac{\sin kL}{k} \approx L - \frac{2iRe(h_1)\delta\beta L^3}{3}$ and $\cos kL \approx 1 - iRe(h_1)\delta\beta L^2$. This allows us to approximate the sum of $A$ and $B$ modes at the end of the periodic medium, (i.e., $L \approx z$), $A + B \propto \frac{1}{\cos kL - (i\delta\beta - h_1)\frac{\sin kL}{k}}$ using Lorentzian function with a quality factor proportional to $\frac{Re(h_1)L^2 + L + \frac{1}{3}Re^2(h_1)L^3}{1 + Re(h_1)L}$. This fits well the Q-factor vs. $L$ simulated using FEM and indicates that Q-factor scales as $\propto L^2$.

## 2. Dependence of the grating DC manifesting an EP (DC$_{EP}$) in the band diagram on the grating groove depth

The detailed FEM analysis indicates the noticeable deviation from the 50 % DC where the EP is observed in a band diagram with the larger grating depth (Fig. S2), and DC$_{EP}$ approaches 50 % for shallow gratings. The results are obtained using FEM simulations of a unit cell. According to Fig. S2, the DC$_{EP}$ grows with the grating groove depths. The EPs appear in the band diagram for an infinite shallow grating of 85 nm nominal groove depth at the specific duty cycle of DC ≈ 0.527 (Fig. 1 e), which is close to but not exactly 50 %. Although $h_2$ vanishes at DC = 50 %[1], for partially etched deep gratings $h_1$ is complex requiring detuning DC from 50 % to compensate for the imaginary part of $h_1$ and achieve the EP condition $Re(h_2 + ih_1) = 0$.

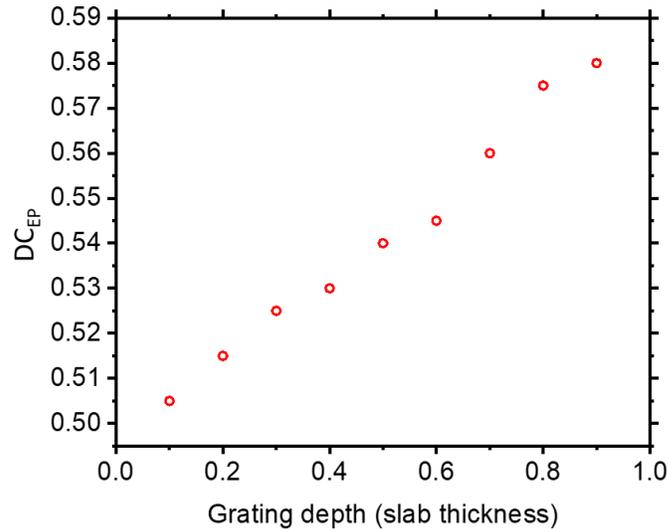

Fig. S2. Dependence of DC corresponding to the EP in a band diagram on the grating groove depth. The grating depth is shown in fractional units of the SiN$_x$ slab thickness.

## 3. Light loss profiles in non-Hermitian periodically structured media using CMT and FDTD simulations

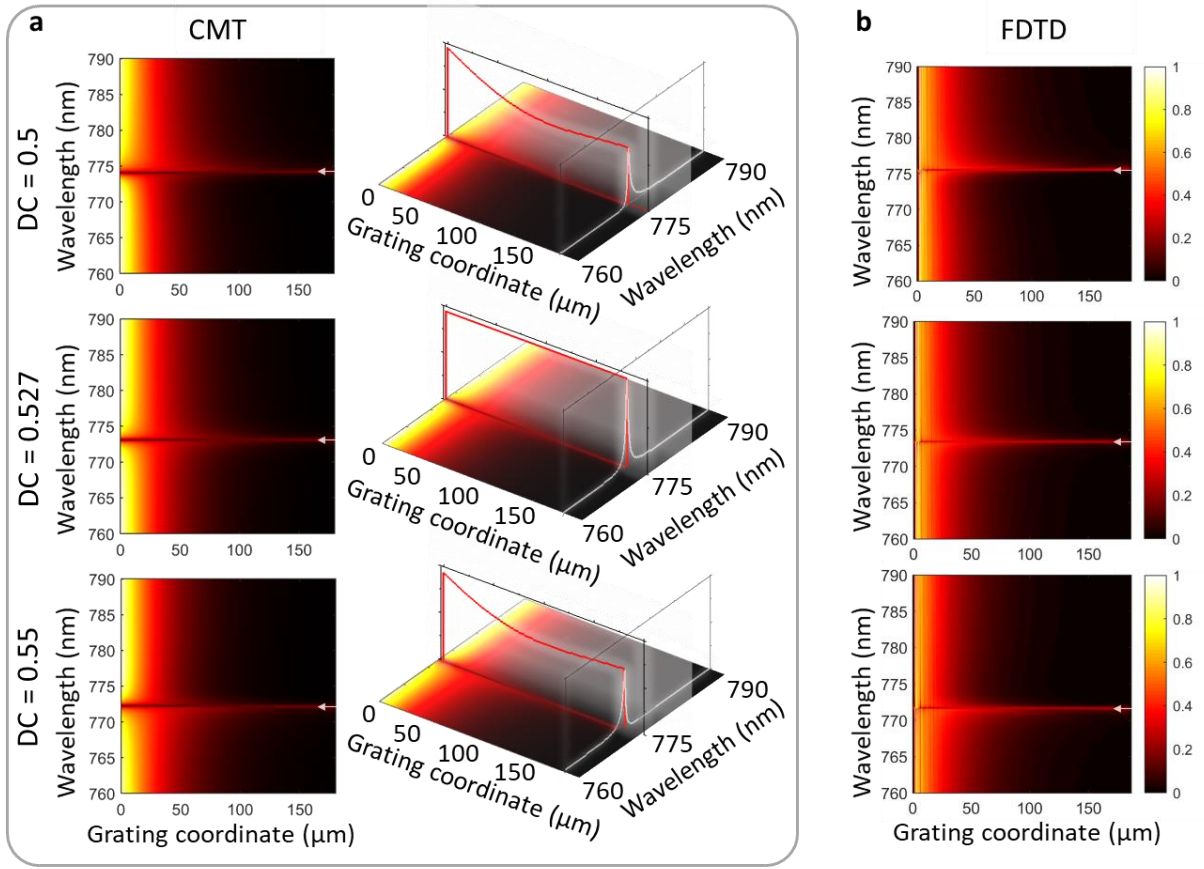

Fig. S3. Light loss profiles in periodically structured media. The normalized amplitudes radiated into free space by a grating waveguide excited from one side ($z = 0$) as a function of wavelength and position $z$, for different grating duty cycles (DC = 0.5, 0.527 and 0.55, top to bottom) obtained using CMT (a) and simulated using FDTD (b). White arrows point out the deep wave penetration. White curves depict the spectra at the grating ends ($z = 180$ μm), and amplitude profiles (red curves) are drawn through the maxima of the corresponding spectra.

## 4. Wavelength dependence of radiated intensity at 50 µm, 100 µm, and 170 µm grating coordinate

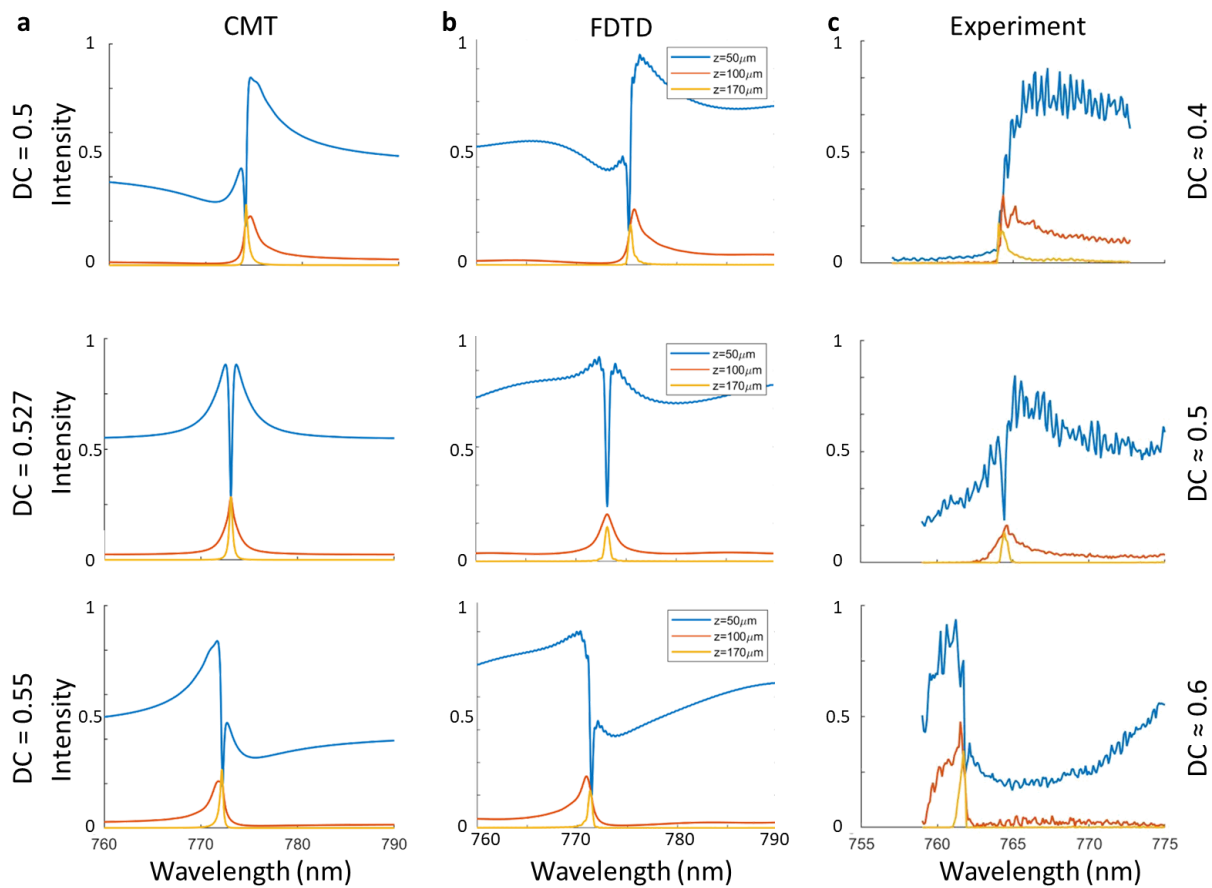

Fig. S4. Free-space radiation spectra of 180 µm long grating at $z$ = 50 µm (blue curve), 100 µm (red curve), and 170 µm (yellow curve) obtained using CMT (a), FDTD (b) for DC = 0.5, 0.527 and 0.55 (top to bottom). (c) Experimentally measured radiated intensities collected from gratings with DC ≈ 0.4, 0.5 and 0.6 (top to bottom).

## 5. Device operation *versus* parameter variation

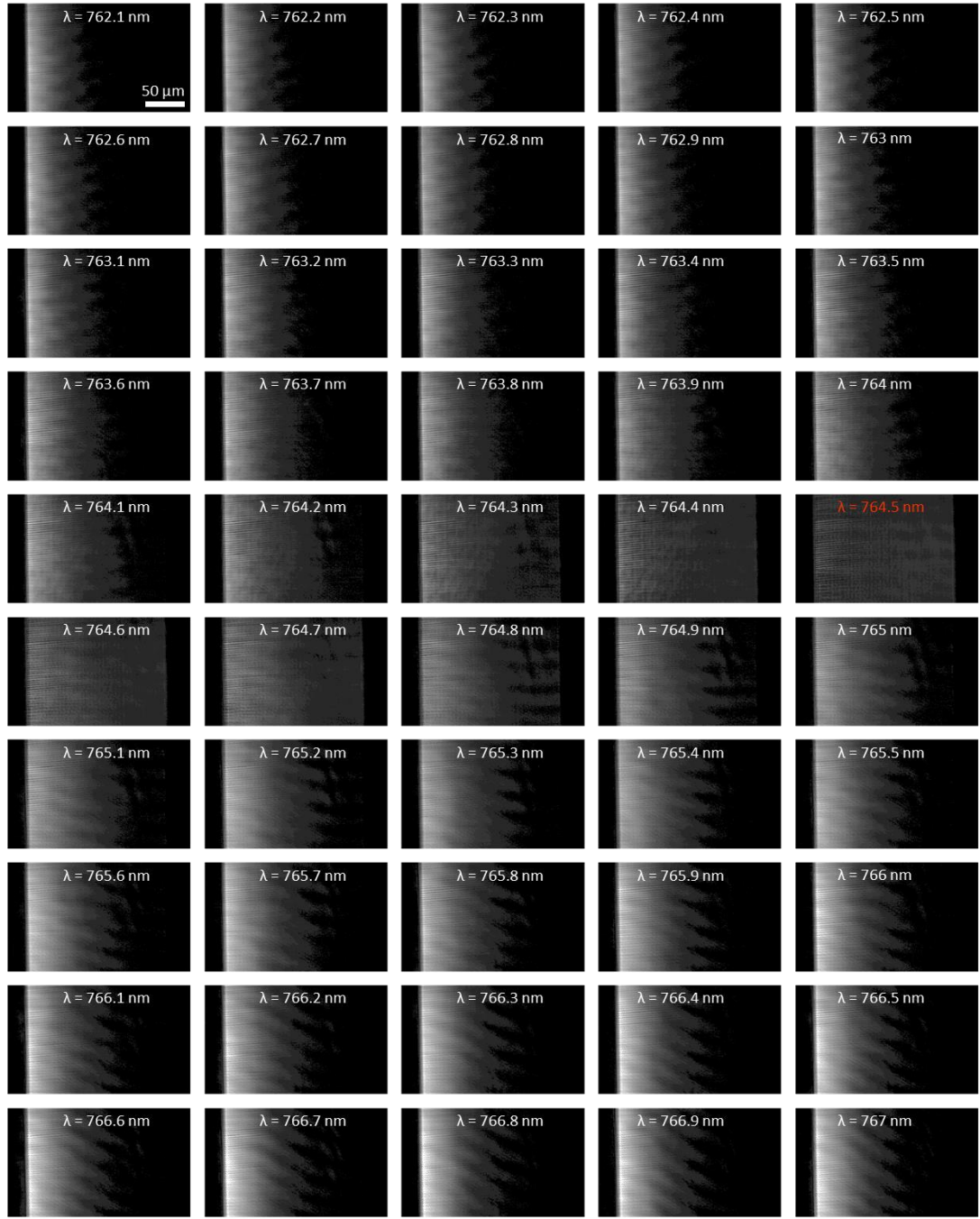

Fig. S5. Top-view optical images depicting intensity profiles of projected beams at different free-space laser wavelengths. Sample 1. The red label indicates tuning into the exceptional point and linear energy loss regime. $L$ = 180 μm. DC ≈ 0.5.

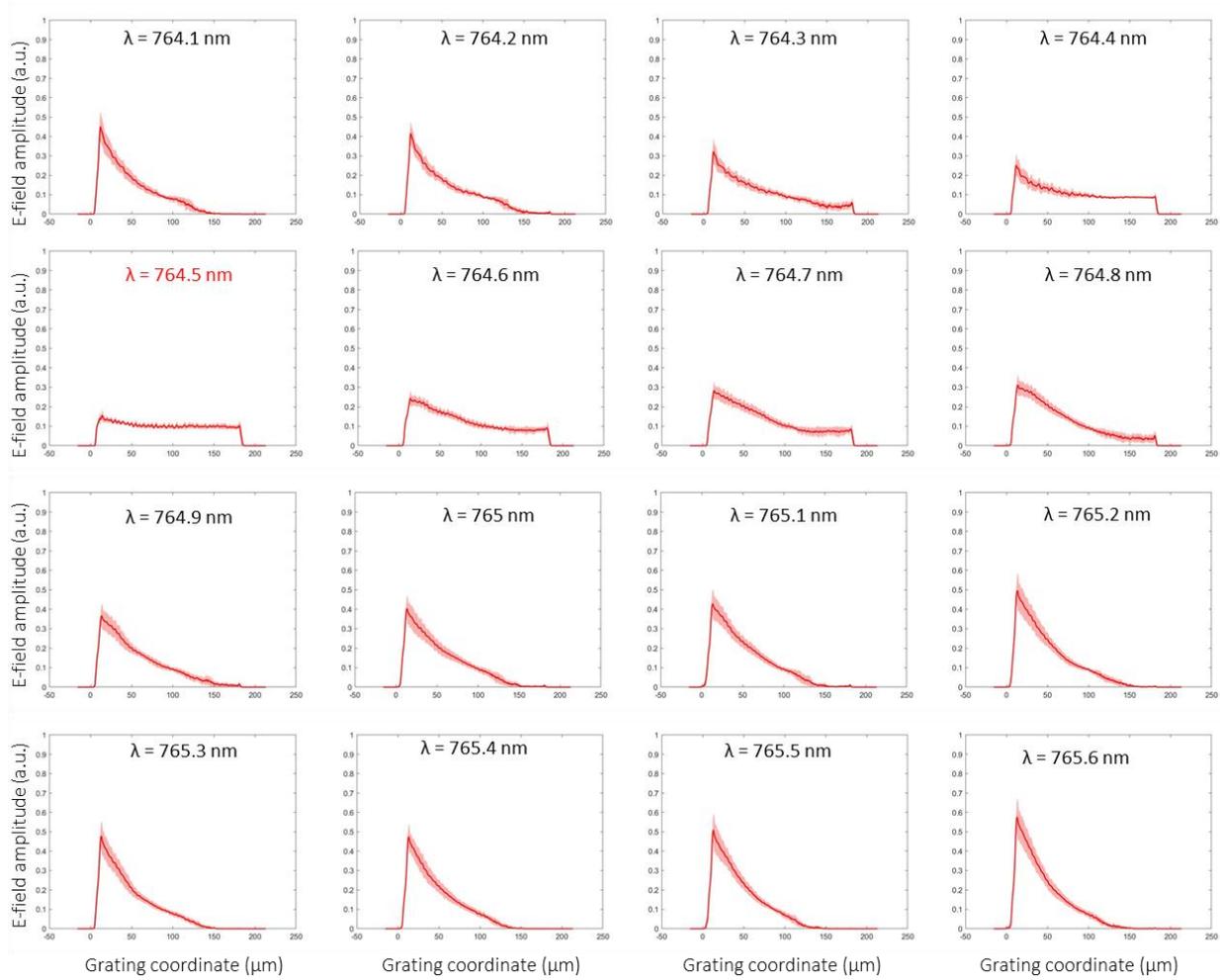

Fig. S6. Spatial free-space profiles vs. excitation wavelength. $L = 180$ μm. DC ≈ 0.5. Sample 1. The curves are obtained by integrating optical images shown in Fig. S5 over the vertical axis. One standard deviation statistical uncertainty shown by the shaded areas is obtained by analyzing multiple separate slices of the image. The red label indicates tuning into the exceptional point and linear energy loss regime.

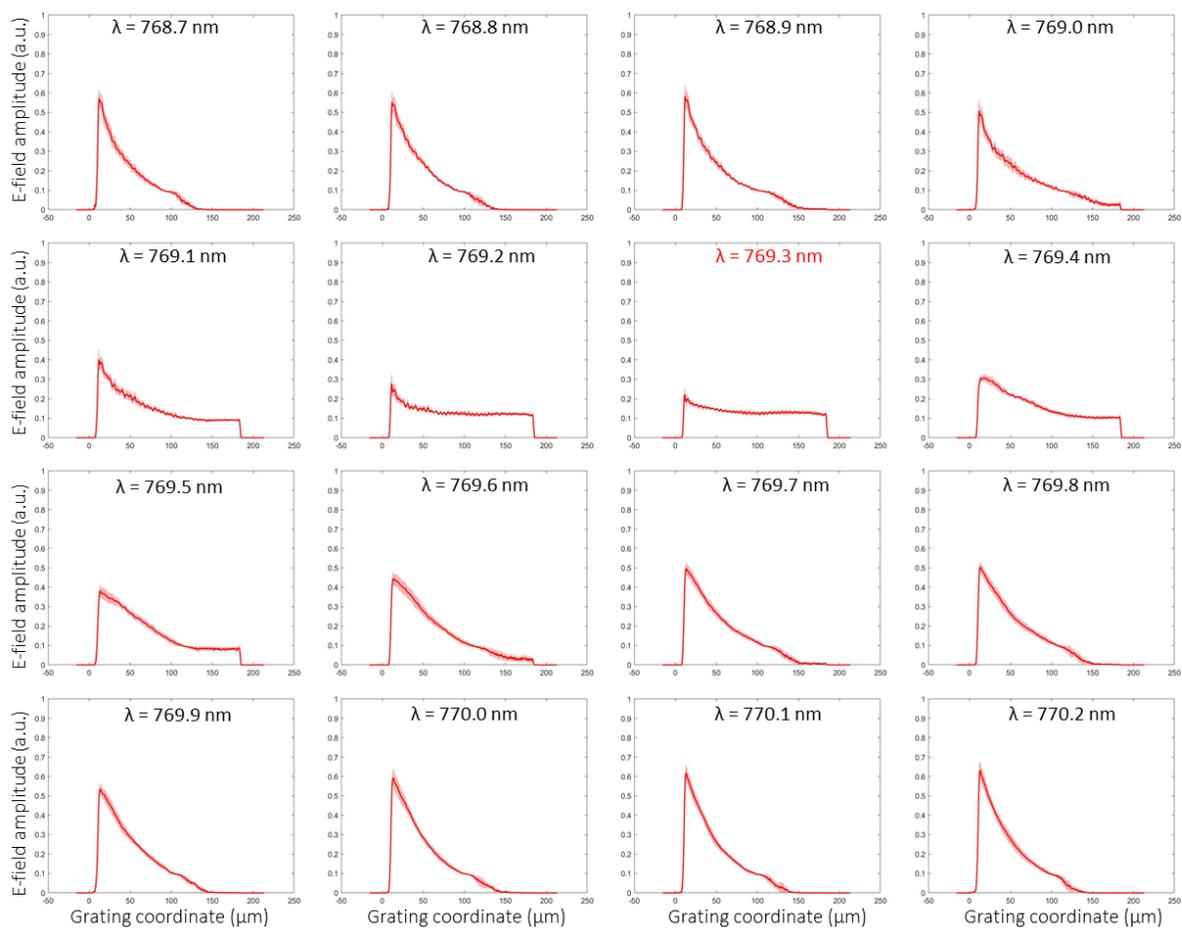

Fig. S7. Spatial free-space profiles vs. excitation wavelength. $L$ = 180 μm. DC ≈ 0.5. Sample 2. One standard deviation statistical uncertainty shown by the shaded areas is obtained by analyzing multiple separate slices of the image. The red label indicates tuning into the exceptional point and linear energy loss regime.

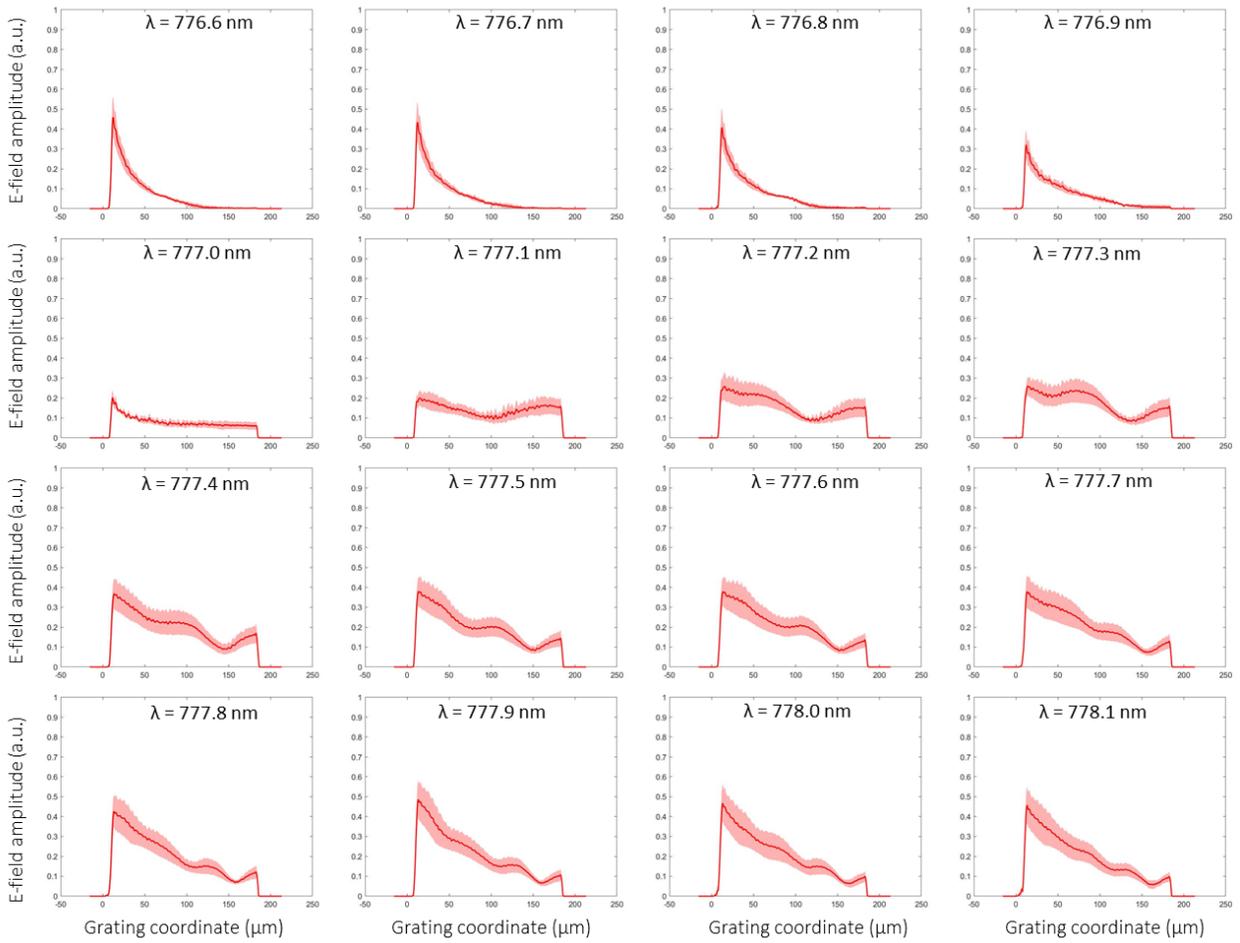

Fig. S8. Spatial free-space profiles vs. excitation wavelength. $L$ = 180 μm. DC ≈ 0.4. One standard deviation statistical uncertainty shown by the shaded areas is obtained by analyzing multiple separate slices of the image.

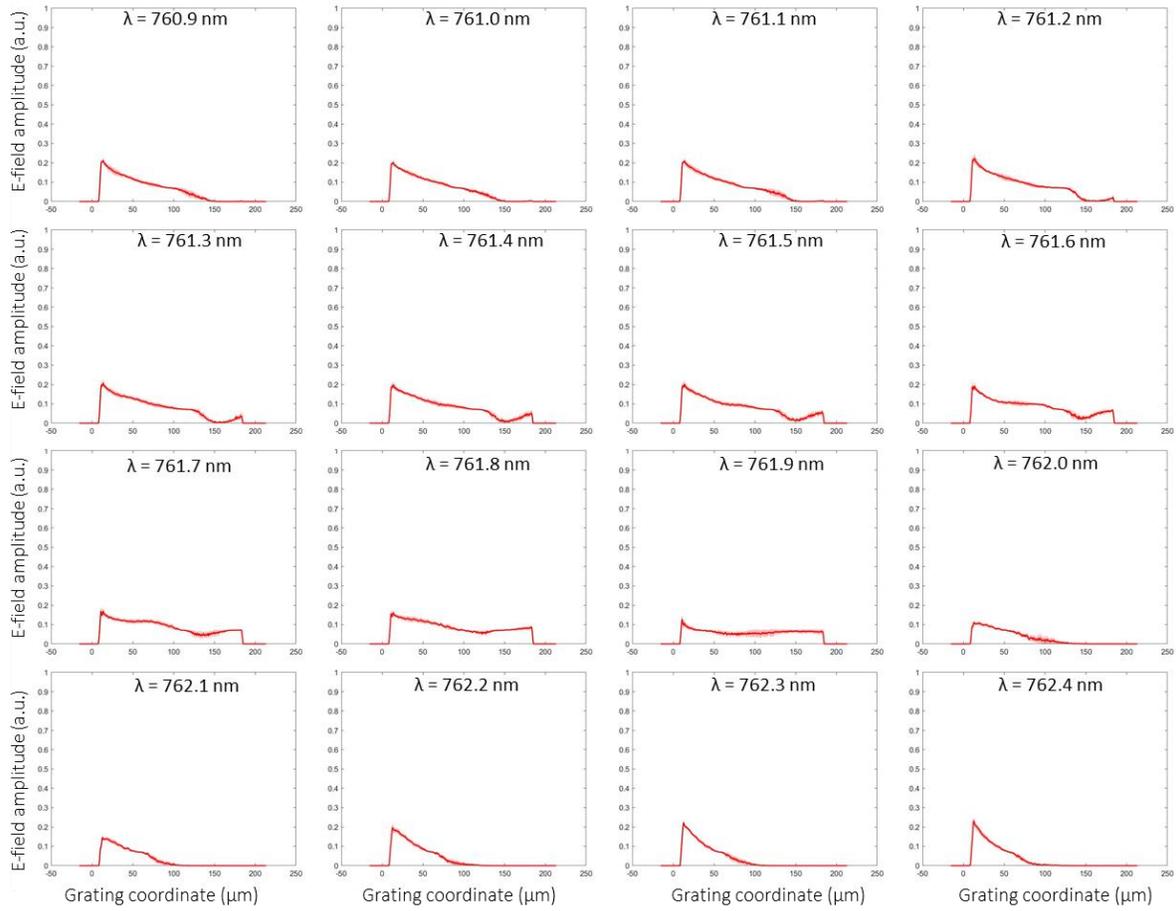

Fig. S9. Spatial free-space profiles vs. excitation wavelength. $L$ = 180 μm. DC ≈ 0.6. One standard deviation statistical uncertainty shown by the shaded areas is obtained by analyzing multiple separate slices of the image.

## 6. The effects of spatial non-uniformity in DC distribution across the grating waveguide

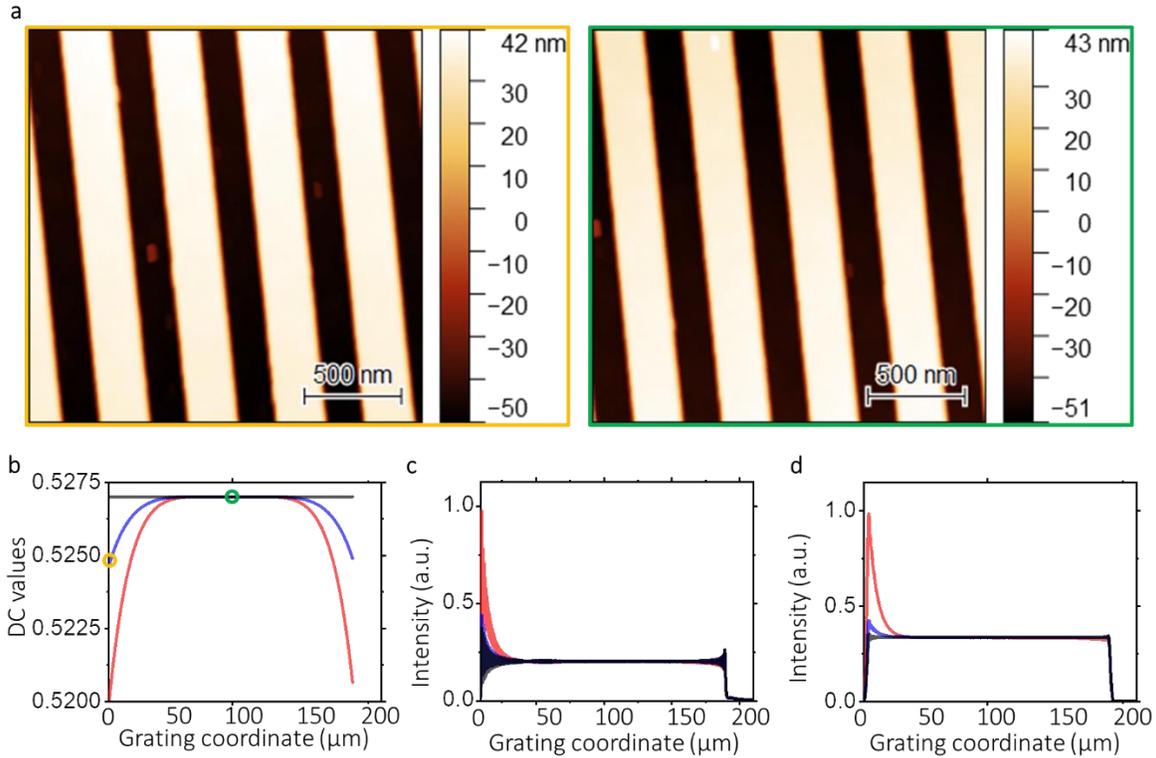

Fig. S10. FEM simulations of the grating performance with DC uniformity affected by an imperfect correction of the electron-beam lithography (EBL) proximity effect. (a) Atomic force microscopy images of the unclad photonic grating (nominal DC = 0.53) at the leading edge (left) and in the grating center (right). The DC at the grating end is ≈ 1 nm smaller than the value at the grating center due to the imperfect correction for the EBL proximity effect. (b) Approximation of the DC variation at the grating ends using 6th order polynomial: $DC(z<L/2) = 0.527 - l/Per*(2z/L)^6$, for $l$ = 0 nm (black curve), 1 nm (blue curve), and 3 nm (red curve). (c) FEM intensity profile of the grating with DC variation in (b). (d) FEM intensity profile convolved with 1.6 μm radius point spread function of the large field of view microscope objective used in the experiment.

## 7. Characterization of the free-space top-hat beam propagation in free space

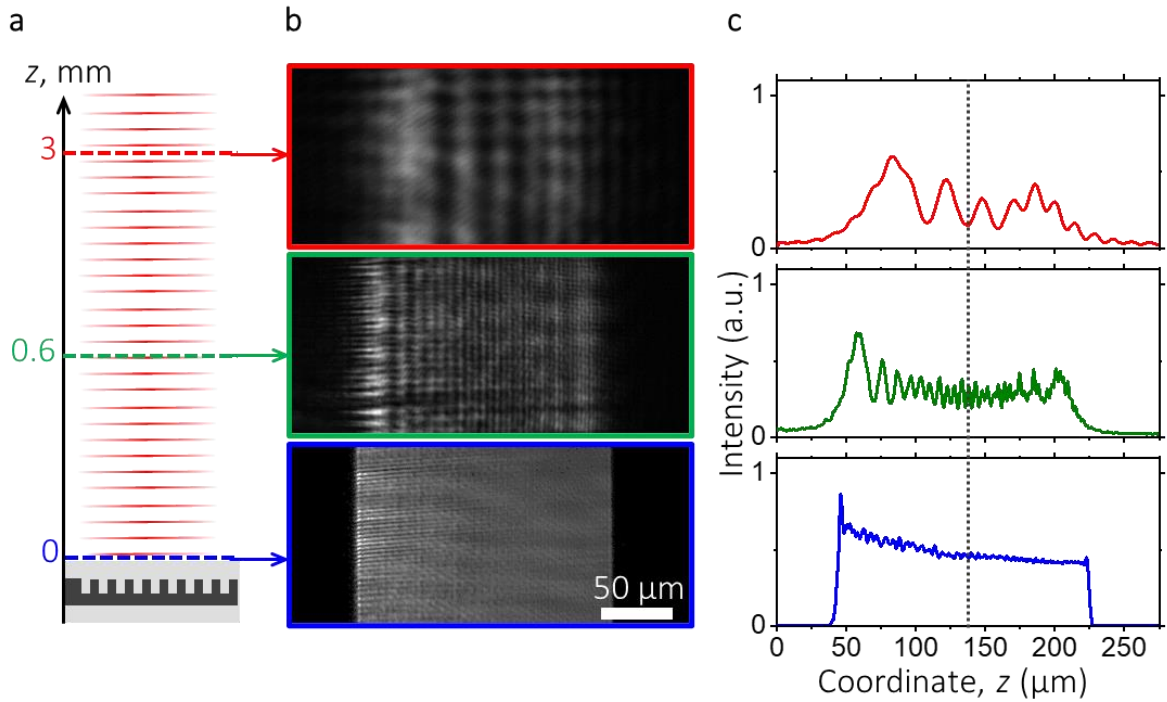

Fig. S11. Beam propagation in free space projected from the photonic grating with a frequency tuned in EP. (a) A schematic depicting a collimated decay-free beam and $z$ coordinates where images of the intensity profiles have been collected. (b) Optical images captured at the chip surface (bottom image), 0.6 mm above the chip (middle image), and 3 mm above the chip (top image). (c) The intensity profiles that correspond to images in panel (b) depicting symmetric beam diffraction once light propagates first 3 mm from the chip.